# Transfer of planar orders onto a sphere: formation and properties of complex topological defects

D. S. Roshal, C.Yu. Petrov, A.E. Myasnikova, and S.B. Rochal

Faculty of Physics, Southern Federal University, 5 Zorge str., 344090 Rostov-on-Don, Russia

General topological principles how to transfer the planar orders onto a sphere are considered. Formation of extended topological defects (ETDs), which have a reconstructed inner structure surrounded by perfect initial order, is discussed. Topological charge of the ETD can be determined from the shape of a characteristic polygon bounding the defect. Relation between the total topological charge of all defects in the spherical structure and the type of initial planar order is found. It is also demonstrated that in the spherical hexagonal crystal a dislocation located in the ETD area is actually absorbed by it, because the order outside the defect doesn't display existence of dislocation in any way. For the case of singly connected spherical hexagonal order arising from mutual repulsion of $N$ particles ($N \leq 1000$) only triangulation of the order inside the ETD regions recovers the linear scars which represent a narrow parts of wider ETD areas.

## 1. Introduction

Two-dimensional (2D) ordered structures with an unusual topology are under discussion since the very beginning of the 20$^{th}$ century. Trying to explain the periodic law of Mendeleev, J. J. Thomson proposed a model of atom, according to which the electrons confined at the sphere surface interact by means of Coulomb potential. Determination of the equilibrium spherical position of repelling equally-charged particles was called the Thomson problem [1]. Later, it was generalized to the case of non-Coulomb potentials [2]. Tammes considered the similar problem how $N$ identical spherical caps should be packed on the sphere to provide the maximal cap size [3]. Experimental investigation of the behavior of colloidal particles located at the interface between two liquids was started by Ramsden [4] in 1903. More than a century later this study led to the synthesis of nanoporous capsules - colloidosoms [5]. Similar ordered structures appear in various systems. For example, they are formed by viral capsid proteins [6, 7], localized electrons in multi-electron bubbles in superfluid helium [8], Pickering emulsion on spherical surfaces [9, 10, 11] and even occur in coding theory [12, 13]. All these natural and synthetic objects are more or less ordered structures forming the 2D closed shells topologically equivalent to a sphere. Due to the curved topology new crystallographic peculiarities appear in these systems [14]. One of such peculiarities is the inevitable existence of topological defects that causes the curvature of 2D crystalline and quasicrystalline spherical structures. Depending on presence of the other types of defects the spherical structures can be divided into two groups.

The first group contains the 'perfect' spherical crystals and quasicrystals. Their structures include the regular topological defects as intrinsic structural components. The other types of defects are usually absent. One of the related examples is presented by a significant number of viral capsids, which have the symmetry of icosahedron rotation group *I* and are described by Caspar and Klug (CK) geometrical model [6]. This capsid model is constructed on the basis of icosahedron net decorated by the periodic hexagonal structure. Asymmetric proteins can occupy the trivial symmetry positions only. Their number $N$ in icosahedral capsids is equal to *60T*, where for CK model $T = h^2 + k^2 + hk$, $h$ and $k$ are integer. Topological defects in these capsids possess 5-fold symmetry and they are located around the 12 vertices of the icosahedron.

Another qualitatively different example of the 'perfect' structures is related to the capsids with the same point symmetry but with the completely different local order type. The pentagonal quasi-crystalline organization of proteins in capsids, which are not described by CK model, is commensurate with the dodecahedron net and the topological defects arising are located around the 20 dodecahedron vertices [15]. Some metal nanoclusters with icosahedral symmetry $I_h$ are arranged like the CK capsids. However, atoms in contrast to proteins can occupy high-symmetry positions forming the Mackay icosahedral shell with faces decorated by a simple hexagonal packing [16]. For this organization of particles, the restriction on their number in the shell is $N = 10T + 2$, and twelve topological defects coincide with the icosahedral nanocluster vertices.

The second group of spherical structures includes less ordered systems, e.g. colloidosomes and structures formed by solid particles from the Pickering emulsion on the spherical surfaces [5, 9, 11]. Despite the fact that these objects are characterized by a more or less perfect hexagonal order, their important features are not so symmetric arrangement of topological defects and presence of other non-topological defects conventional for 2D planar lattices. One of the first pioneer works **[9]** devoted to the peculiarities of the spherical hexagonal order was published a decade ago. It was found that this order type is very sensitive to ratio $R/a$, where $R$ is the radius of the sphere and $a$ is an average particle radius. For $R/a \geq 5$ the authors of Ref. [10] have found linear defects, which they called grain boundaries, or scars. In fact, these defects were the chains consisting of closely located particles with different surroundings. Particles having 5 or 7 nearest neighbors sequentially alternate in the chains. Less ordered spherical structures demonstrate also the other defects unconventional for the planar geometry. Recently, formation of ETDs with a square order inside, which are located on the colloidosome surface [5], was explained [17]. Since the particle number $N$ can vary at constant $R/a$ ratio, the average packing density can be changed, accordingly. If the number $N$ approaches to the minimum possible value (at smaller $N$ the sphere surface will not be completely covered), the excesses of the sphere area per one colloidal particle can aggregate in one place, where the ETD with a square order inside is formed [17].

Essential order peculiarities in the spherical objects from the both groups can be understood on the basis of construction and investigation of planar nets of their structures [6, 15, 17]. The general aim of this work is to continue the study of principles how the planar order is transferred onto a sphere. In the framework of the theory developed we demonstrate some universal characteristics and properties of defects arising due to this transfer and generalize the notion of topological charge for the orders, triangulation of which doesn't have any physics sense. Besides, we relate the total topological charges of all defects in the spherical structure with the type of initial planar order. We also discuss the interaction between topological defects and dislocations in the spherical hexagonal order and come to the surprising conclusion that a complex object representing a superposition of topological defect with a minimal total charge and dislocation is impossible.

The paper is organized as follows. In the next section, we discuss the relations between the Euler's theorem, topological charges and planar nets of some spherical structures. The third section demonstrates the absorption of dislocations by the topological defects. The last section is devoted to the discussion of our results.

## 2. Relation between the Euler's theorem, topological charges and planar nets of spherical structures

Following [14], some peculiarities of the point defect formation in the hexagonal order on the sphere can be understood on the basis of Euler theorem [18]. According to it for the polyhedron that is topologically equivalent to the sphere the relation between the number of vertices $V$, the number of edges $E$, and the number of faces $F$ reads

$$V - E + F = 2.  \qquad (1)$$

To apply Eq. (2) to spherical structures the neighboring particles are connected to form a polyhedron with triangular faces. Any algorithm (see for example [19]) can be used for the triangulation.

Since each face is bounded by three edges and each edge adjoins to two faces, we have

$$F = \frac{2}{3}E. \qquad (2)$$

Let the number of vertices from which $n$ edges originate be $V_n$, and the number of edge ends originated from such vertices be $E_n$. Substitution of Eq. (2) into Eq. (1) results in

$$6V - \sum_n E_n = 12, \qquad (3)$$

where $E_n = nV_n$, and $V = \sum_n V_n$. Simplification of Eq. (3) yields

$$3V_3 + 2V_4 + V_5 - V_7 - 2V_8 - \ldots = 12. \qquad (4)$$

The coefficient $q_n$ at $V_n$ is usually noted as the topological charge of the $n$-angle defect. For example, the topological charges of pentagonal and quadrangular defects are +1 and +2, respectively. Thus, the sum of topological charges of all point defects on the sphere surface is equal to 12.

Here we propose to examine the topological features of the defects formation in a different and more general way not based on the full triangulation of the spherical structure. Instead we study how to transfer the planar order of different types (e. g. quasicrystalline order) onto the sphere using an appropriate closed polyhedron. Particular decorations of the polyhedron faces by the initial planar order allow subsequent smoothly mapping the order onto a sphere, and the topological defects appear obligatory in the polyhedron vertices only. To obtain the polyhedron net the sectors are cut out or inserted into the initial planar structure. The minimal angular value of the appropriate sectors is determined by the order type. For example, for the well known hexagonal lattice the smallest value of the sector angle is equal to $\pi/3$ and the defects with topological charge of 1 and -1 correspond to the eliminated and inserted smallest sectors. A greater sector should have the angle value multiple to $\pi/3$. The sector with an angle of $\pi/2$ can be cut out, or inserted into a square lattice. Another, less trivial example is presented by decagonal and pentagonal Penrose tilings [20, 21]. Both these structures have ten-fold rotational axes. Therefore, the unit positive topological charge corresponds to the eliminated sector with the angular value $\pi/5$ and a more symmetrical way to transfer these types of quasicrystalline order to the sphere is to use a net of dodecahedron. An example of the dodecahedron net decorated by the ordinary Penrose tiling is demonstrated in Fig. 1. One more example of mapping a chiral pentagonal order onto a sphere can be found in [15].

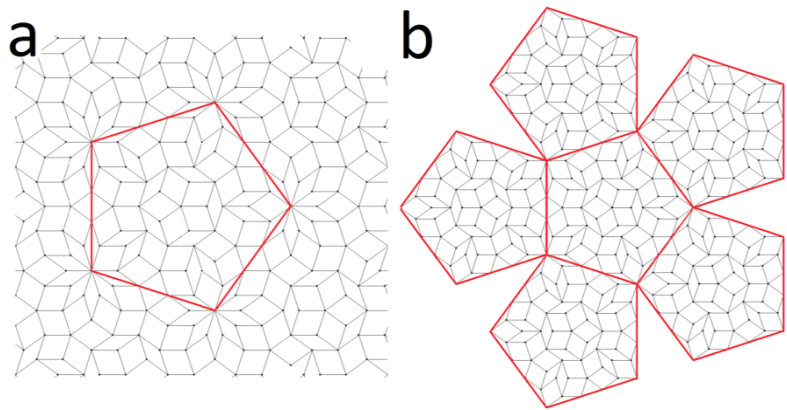

FIG. 1 (color online). High-symmetry decoration of the dodecahedron net with Penrose tiling: (a) Conventional Penrose tiling with the global 5-fold axis located in the panel center. Big pentagon presents the dodecahedron face. Its vertices are located in the approximate 10-fold axis of the tiling. (b) A part of the dodecahedron net decorated with Penrose tiling. The angular value of the eliminated sectors is equal to $\pi/5$. Such a cut out sector generates the unit positive topological charge for this type of order.

Since the unit topological charge is simply determined by the geometry of the order transferred we can easily calculate the number and type of topological defects that should appear during the transfer. To perform this calculation let us recall some geometrical properties of a closed polyhedron which is topologically equivalent to a sphere and satisfies Eq. (1). It is simpler to start from the polyhedron with triangular faces. In this case the sum of all planar angles of all its faces is obviously equal to $\pi F$. Application of Eqs. (1-3) for polyhedron with triangular faces yields that $\pi F = 2\pi V - 4\pi$ or in other words the algebraic sum of angles of all the cut out and inserted sectors in the polyhedron net is equal to $4\pi$.

This conclusion is easily generalized for polyhedron with arbitrary planar faces since a planar face with $n$ edges can be constructed from $n$ triangles with a common vertices located at this face. Therefore for all polyhedrons with planar faces the algebraic sum of angles of all cut out and inserted sectors in their nets is also equal to $4\pi$. This statement can be used to calculate the total topological charge of all defects in spherical structures originating from planar orders of different types. Indeed, let $\Omega$ denotes the angular value of the smallest cut out sector for the order transferred. Let us assign to the minimal eliminated or inserted sector the topological charge equal to 1 or -1, respectively. Then for the spherical structure with this type of order the total topological charge of all defects is equal to $4\pi / \Omega$.

Note that while the sector is cut out or inserted the area near the top of the resulting solid angle can be arbitrary rearranged. The initial planar order can disappear completely within the ETD area. But this reconstruction cannot change the total topological charge of the defect provided it is completely surrounded by the perfect initial planar order. It is useful to surround the defect with the characteristic polygon, which satisfies two following conditions: 1) the sides of this polygon pass through the order nodes; 2) the angle between the nearest polygon sides in the initial planar order is equal to $\pi - 2\pi/N$. Here $N$ characterizes the rotational symmetry of the initial order, i.e. $N=4$ and $N=6$ for square and hexagonal lattices, accordingly, while for the Penrose tiling $N=10$. Then the total topological charge $q$ of the defect is simply defined as
$$q = N - m , \qquad (5)$$
where $m$ is the number of sides of the characteristic polygon. The conventional dislocation without a topological charge is always surrounded by the characteristic polygon with $N$ sides. In the well-known hexagonal case $N=6$ and the topological defects with the unit positive charge are surrounded by pentagons (see Fig. 2).

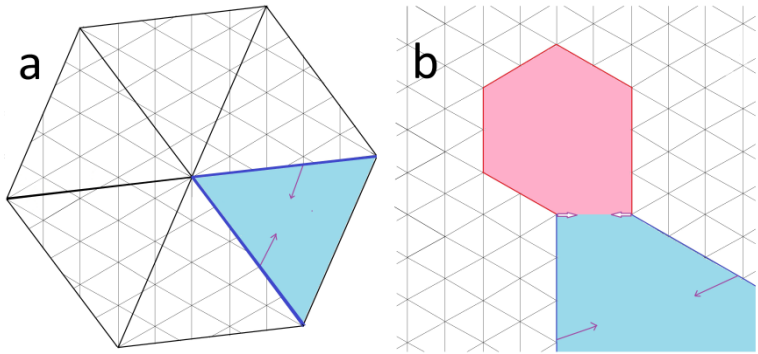

FIG. 2 (color online). Nets of two topological defects arising due to the transfer of the hexagonal lattice onto the spherical surface. In the both cases the value of the eliminated sectors (shown in blue) is equal to $\pi/3$ and the both defects have the same unit topological charge. In the spherical structure the defects are surrounded by characteristic pentagons, whose sides are parallel to the minimal translations of the initial hexagonal order. (a) The ordinary point topological defect. (b) The extended topological defect bounded by the pentagon with not equal sides.

However, even for the simplest case shown in Fig. 2 the sector with the unit topological charge can be cut out in essentially different ways. Besides the simplest topological defect (shown in Fig. 2(a)) a lot of ETDs with different nets is possible. The net of one of these defects is shown in Fig. 2(b). The pentagon surrounding the defect has one side longer than the others. This fact creates the impression that line of extra nodes passes through the longer side of the pentagon and ends by a dislocation in the defect area. However, it is no so and the precise analysis of the order outside the defect area proves this fact. The next section explains why the ETD absorbs dislocations located inside its area.

## 3. Absorption of dislocations by topological defects in the spherical hexagonal order

Characteristic polygon surrounding the topological defect determines its total topological charge like the Burgers contour specifies the conventional dislocation.

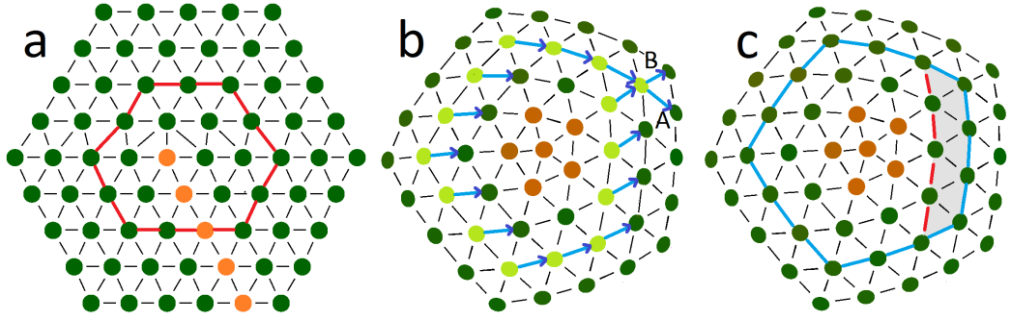

FIG. 3 (color online). Simple defects of the hexagonal order. Both the dislocation (panel (a)) and the topological pentagonal defect (panel (b), (c)) are surrounded by polygons; outside the polygons the local translational properties of the lattice are unchanged. Panel (b) shows how the basic translation **A** turns into the other basic translation **B** after a circulation around the topological defect along an arbitrary chain of nearest nodes (shown in light-green color). The angle Θ between the vectors **A** and **B** determines the topological charge of the defect. Panel (c) demonstrates how the characteristic pentagon is changed depending on the case whether the node line is cut or inserted.

Figure 3 shows the simplest dislocation (panel (a)) and the point topological defect ($q_5 = 1$) of the hexagonal order. Both the defects can be surrounded by the characteristic polygons (cases (a) and (c)) or by an arbitrary contour (case (b)). Anyway outside the defect area the *local* translational properties of the lattice are conserved. It is simpler to determine the topological charge from the shape of characteristic polygon, but the integral characteristic of any arbitrary contour bounding the defect also allows its charge determination [14]. Such a contour presents a polygon formed by a closed chain of nearest nodes related to the undistorted order. Let us consider the following sum over this chain:

$$\sum_i \Delta \mathbf{A}_i , \qquad (6)$$

where *i* is the node number, **A** is one of the basis translations (or quasitranslations) and $\Delta \mathbf{A}_i$ is its deviation associated with the node (*i*). At motion along the chain (and around the defect) the direction of the vector $\mathbf{A}_i$ is changed slightly. After the circulation around the dislocation the vector A coincides with itself and sum (6) is zero. In contrast, after going around the topological defect the translation **A** coincides with the other translation **B**, which is rotationally equivalent to A, and the sum (6) is simplified to **B**-**A**. The angle between the vectors **A** and B corresponds to the value of removed (see Fig. 2(b)) or inserted sector that, in turn, determines the value of the topological charge of the defect. Note also that all the local lattice translations during the circulation around the contour are rotated simultaneously and the angle Θ of their total rotation is completely independent on the initial choice of the vector **A** direction. Finally, for the order characterized by *N*-fold rotational axes the topological charge of the ETD is determined as

$$q = \frac{N\Theta}{2\pi} \qquad (7)$$

Since for the contour surrounding the dislocation the sum (6) is equal to zero, this contour can be translated while the dislocation is located within it. The contour surrounding the topological defect is not translationally invariant; any translation results in its breaking.

However, in the particular case of hexagonal lattice the characteristic pentagon surrounding the topological defect can be changed by adding or cutting strips, parallel to its sides (see Fig. 3(c)). For example, after cutting a strip the pentagon side parallel to the strip becomes one node longer, and each of two adjacent sides becomes one node shorter. As a result, the contour perimeter is decreased by one node. Analogously, one can reduce the scalene hexagonal Burgers contour surrounding the conventional dislocation shown in fig. 3 (a). This contour has the well-known invariant characteristic which is the Burgers vector. Therefore the following question arises. Is it possible to state that the dislocation coexists with the topological defect provided the defect area is surrounded by a particular scalene characteristic pentagon and is it possible to find an analog of the Burgers vector for such a case?

Let us formulate the problem using a more rigorous mathematical language. The pentagon surrounding the topological defect can be characterized by a five-dimensional integer vector **S**={$n_1$, $n_2$, $n_3$, $n_4$, $n_5$}, its components are equal to the lengths of the pentagon sides. Cutting the strip parallel to first side of the contour is equivalent to adding the translation

$$\mathbf{a_1} = <1,-1, 0, 0,-1> \qquad (8)$$

to the vector **S**. The cutting operations for four other sides correspond to translations obtained from vector (8) by the cyclic permutation of its components. Adding of stripes corresponds to the same translations, taken with the opposite sign. Let the symmetric matrix $M_{ij}$ consists of lines $a_i$. Then

$$\Delta S_i = M_{ij} V_j , \qquad (9)$$

where the component $\Delta S_i$ is equal to the length change of the side with the number *i*, and component $V_j$ specifies the number of side *j* cutting. The negative sign of the vector component $V_j$ means that the corresponding side increases.

Matrix $M_{ij}^{-1}$ entering the opposite relationship

$$V_i = M_{ij}^{-1} \Delta S_j \qquad (10)$$

is integer:

$$M_{ij}^{-1} = \begin{bmatrix} 1 & 0 & -1 & -1 & 0 \\ 0 & 1 & 0 & -1 & -1 \\ -1 & 0 & 1 & 0 & -1 \\ -1 & -1 & 0 & 1 & 0 \\ 0 & -1 & -1 & 0 & 1 \end{bmatrix} \qquad (11)$$

The rather surprising fact that all coefficients of matrix (11) are integer means that any pentagon surrounding the topological defect can be transformed into the equilateral pentagon by adding strips operations.

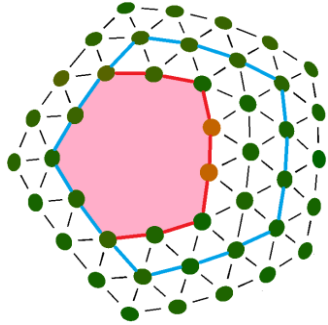

FIG. 4 (color online). Absorption of dislocation by the topological defect. At first glance it seems that an excess line of nodes crosses the longer side of the internal pentagon and ends by the dislocation inside the contour. However, as it is demonstrated in previous discussion, the internal contour can always be surrounded by the outer equilateral pentagon. The order outside the outer pentagon is obviously perfect and no sign of excess line of nodes can be found.

For example, applying a set of operations **V**=<0,-1,-2,-1, 0> to the irregular pentagon the net of which is shown in Fig. 2(b) one can surround it by an external equilateral pentagon with the side length equal to three (see Fig. 4). From a physical point of view this means that a dislocation that seems to be entering the topological defect is absorbed by it, since the order outside the defect area does not display its existence. The same consideration can be applied not only to the unit positive topological defect, but also to the defect with the unit negative topological charge. The related matrix (analogous to the matrix (11)) describing the variation of heptagon sides consists also of integer coefficients and a similar consideration proves that the defect with a unit negative topological charge also absorbs the dislocations located inside its area.

The above mathematics can be used for a more detailed analysis of ETDs arising in the spherical hexagonal order. With its help one can determine the ETD center and to localize therein the total topological charge of the defect. For this purpose the initial hexagonal order is restored inside the ETD and the surrounding contour around it is constricted to the smallest possible characteristic pentagon. Besides, the ETD can be characterized by the degree of the order violation $\eta = (N^i - N^D)/N^i$, where $N^i$ and $N^D$ are the number of nodes after the order restoration inside the contour is shown in Fig. 5. First, the defect is surrounded by the characteristic pentagon. Then the number of restored strips from each side of the pentagon can be calculated using Eq. (9) : $\Delta S_i = n^0 - n_i$, where $n_i$ is the initial length of the pentagon side. Here we take $n^0 = 1$ that corresponds to the smallest possible final pentagon.

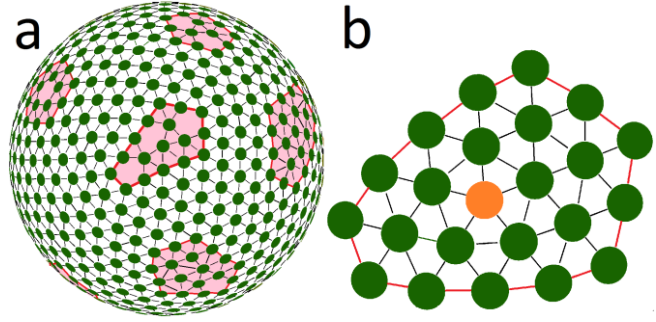

FIG. 5 (color online). Reconstruction of the hexagonal order within the defect area and determination of the defect center. The defect under consideration is located in the center of the panel (a), which shows a model hexagonal order on the spherical surface (see more details in Appendix). Extended defects with the unit topological charge are highlighted as red pentagons. Panel (b) demonstrates the restored order within the defect area.

4. **Discussion and conclusion**

Our method to transfer the planar order of different types onto a sphere using an appropriate closed polyhedron is suitable for different kinds of spherical order and we expect to develop the above ideas in the future publications. But here we would like to discuss mainly the relation between the ETDs appearing in our approach and the scars observed in the spherical hexagonal order and widely discussed in the literature [10, 11, 22].

For that purpose we model the spherical hexagonal order and its defect using the conventional minimization of the free energy based on Lennard-Jones pair potential. For the sake of simplicity we assume that the repulsion between the particles dominates over their attraction. This condition leads to the more uniform distribution of particles on the sphere surface and the appearance of defects in the structure is mainly caused by the spherical topology. (See more details about the model and its simulations in the next section.). Analogously, the spherical hexagonal order was modeled in previous works [11,14, 17] and the structures resulted were similar to the experimental ones.

We have obtained about 50 spherical structures, with the number of ordered repulsive particles from 700 to 1000. In all the cases, the simulated hexagonal order was global. We have found no defects which represent the true grain boundaries and do not allow the continuous circulation around. All the defects found are surrounded by the singly connected hexagonal order. This order corresponds always to more or less defective mapping of a single planar hexagonal order onto the sphere by means of the icosahedron net. We were always able to localize exactly twelve ETDs with the unit positive topological charge. These defects repel each other and are located approximately near the vertices of an icosahedron. A similar icosahedral arrangement of the defects was discussed or observed in the theoretical and experimental works [23, 24, 11]. In contrast, distribution of the conventional dislocations obtained in our study was rather random than regular. We've never found an ETD surrounded by heptagon. The consideration of the related planar net makes this result quite clear. To obtain such a topological defect one should insert the sector into the planar hexagonal order, which creates inevitably an extended area with negative Gaussian curvature near the top of the added sector. Thus this ETD is incompatible with the spherical geometry and its appearance can be expected on the more complex surfaces, in regions where the surface has a saddle shape.

The structure shown in Fig. 5(a) and containing 700 particles is a typical one for our simulations. The ETD located in its center like the most of other defects simulated is rather extended than linear. The nodes in this defect as well as in many other ones form almost perfect small pentagons with empty center. The initial hexagonal order is practically destroyed inside the defect area and due to this fact its triangulation seems to us not completely reasonable procedure. But the triangulation is the only way to recover the scars inside the ETDs (see Fig. 6).

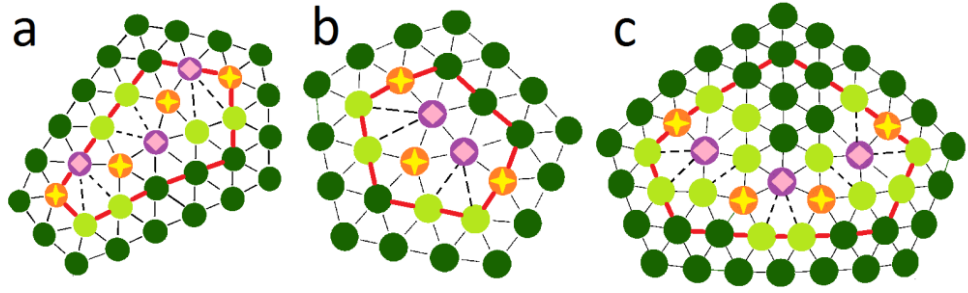

FIG. 6 (color online). Triangulation of the order inside the extended defect region allows recovering the scars. Three extended defects are shown. The nodes with 5 and 7 neighbors form the scars and they are colored in orange and violet. The light-green nodes according to Delaunay triangulation have six neighbors but are not related to the outside hexagonal order. If these nodes are considered the defects take rather extended than linear form. Panels (a-b) present two ETDs of the spherical structure shown in Fig 5(a), while the panel (c) demonstrates the ETD of the other spherical structure formed by 1000 particles.

The scars were initially defined as 'high-angle ($30^o$) grain boundaries, which terminate freely within the crystal' [10]. Let analyze this definition in details. First it is not correct to characterize these defects as high-angle ($30^o$) grain boundaries. In reality, during the circulation along the contour surrounding the ETD the local basis orientation is continuously rotated like it was considered in the previous section. After the complete circuit the total angle of rotation becomes equal to $60^o$. The triangulation of the defect area unreasonably converts the extended defect into the linear one and makes a wrong impression that the continuous $60^o$ rotation is divided into two $30^o$ sub-rotations occurring abruptly at the ends of scar. Second, let us stress that one can discuss the local order orientation only in the region outside the ETD area. Inside the ETD area the formal triangulation assigns 6 neighbors to the nodes which are not related obviously to the hexagonal order. Therefore the nodes which have 5 and 7 neighbors and form the linear scar present only the part of the extended defect area.

Finally, general topological relationships between the planar structures and related spherical orders were investigated and the appearance of the ETDs due to the transfer of the planar order onto the sphere was discussed. The interaction between dislocations and ETDs was considered and the mechanism of dislocation absorption by extended topological defect was proposed. The results presented may be interesting for a wide community of scientists investigating different ordered spherical structures ranging from viral capsids and buckyballs to colloidosomes.

## 5. Appendix. The 2D hexagonal structure self-assembly on the sphere surface

Self-assembly of 2D structure on a non-planar surface can be described by the conditional minimization of the system free energy $F$ with respect to coordinates of the system particles. The condition imposed is that any particle during minimization should be on the surface under consideration. Lennard - Jones potential is one of the simplest potentials, leading to perfect hexagonal order on the plane. That is why it is reasonable to use with certain restrictions this potential for simulation of the hexagonal order on curved surfaces. The free energy of the model system presents the sum of energies of pair interactions of particles, described by Lennard-Jones potential:

$$F = \varepsilon \sum_{j>i}^{N}\left(\left(\frac{\sigma}{r_{ij}}\right)^{12} - 2\left(\frac{\sigma}{r_{ij}}\right)^{6}\right), \quad (12)$$

where $r_{ij}$ is the distance between $i^{th}$ and $j^{th}$ particles, $N$ is the number of particles, $\sigma$ is the distance between the particles in pair, corresponding to the minimum of Lennard-Jones potential.

The structure arising on the sphere surface is determined by the initial particle positions and two parameters: the number of particles $N$ and the ratio $\xi = \sigma / R$, where $R$ is the sphere radius. One can characterize the order under study by the effective distance $r_{eff}$ between neighboring particles, which we define as

$$r_{eff} = a_0 R \sqrt{\frac{1}{N}}, \qquad (13)$$

where geometrical factor $a_0 = \sqrt{\frac{8\pi}{\sqrt{3}}} \approx 3.81$. The introduced quantity (13) is a lower estimation of the average distance between the neighboring particles in the spherical structures. This estimation becomes exact only for 2D hexagonal packing with the same surface particle density $N/4\pi R^2$ as on the sphere. If $\sigma$ is a few percent less than $r_{eff}$, attraction dominates in Lennard - Jones potential and the numerical simulation of the system behavior shows that the particles do not completely cover the sphere surface, but aggregate into several clusters. At somewhat larger value $\sigma$, emerging global single hexagonal order may be strongly distorted and, in particular, the defective area with a square order may appear on the sphere surface [17].

In the simulations of this paper we suggest for the simplicity that $\sigma \gg r_{eff}$. This condition leads to the most uniform distribution of particles on the sphere surface, since the term associated with the repulsion of the particles dominates in Lennard - Jones potential, and the appearance of defects in the structure is caused mainly by the spherical topology. If the number of particles $N$ is sufficiently large ($N \geq 200$-$250$), then the approximately uniform hexagonal order with bounded defects is formed on the sphere. Note also that in the result of the conditional minimization of Eq. (12) (or of any similar repulsion potential) we obtain different equilibrium structures corresponding to the same values of $N$ and $\sigma$ depending on the initial distribution of particles. The equilibrium energies of all these structures are very close [14]. For example, for the number of particle $N \approx 300$ the minima of Eq. (12) differ one from the others by less than a half of percent. If the value of $N$ increases the difference between the equilibrium energies is reduced, however the number of hexagonal structures with the similar energies (which differ from each other by arrangement of defects) grows.

## 6. Acknowledgments

Authors are grateful to Yu. M. Gufan and V.B. Shirokov for fruitful discussions and acknowledge financial support of the RFBR grant 13-02-12085 ofi_m.